\documentclass[conference]{IEEEtran}
\usepackage{flushend}

\ifCLASSINFOpdf
\else
\fi
\usepackage[cmex10]{amsmath}
\usepackage{algorithm}
\usepackage{algorithmic}
\usepackage[tight,footnotesize]{subfigure}
\usepackage{graphicx}
\begin{document}
%
\title{DRUG: An Energy-Efficient Data-Centric Routing Protocol for Wireless Sensor Networks}

\author{\IEEEauthorblockN{B. P. S. Sahoo}
\IEEEauthorblockA{Dept. of Computer Science \& Engineering\\
Indian Institute of Technology Kanpur, India\\
Email: bsahoo@iitk.ac.in}
\and
\IEEEauthorblockN{Deepak Puthal}
\IEEEauthorblockA{Qatar Mobility Innovations Centre (QMIC)\\
Qatar Science and Technology Park, Doha, Qatar.\\
Email: deepakp@qmic.com}
}


%


\maketitle

\begin{abstract}
In general, sensor nodes are deployed in left unattended area. In such situation feeding energy to the batteries or replacing the batteries is difficult or even sometimes impossible too. Therefore, prolonging the network lifetime is an important optimization goal in this aspect. In this paper, we propose a new Energy-efficient Datacentric RoUtinG protocol called DRUG. In this paper, we propose an adaptive Data centric approach to find an optimal routing path from source to sink when the sensor nodes are deployed randomly deployed in a restricted service area with single sink. Using the NS-2 Simulator, we compare the performance of DRUG with that of the FLOODING and SPIN protocol.
\end{abstract}



%
\IEEEpeerreviewmaketitle

\section{Introduction}
In recent times, the increase in the low cost sensor node design and flexibity has led to an increase in the development of and the demand for rich wireless sensor network (WSN) applications[1, 2]. Efficient design and implementation of WSNs has become a hot area of research in recent years. By networking a large numbers of tiny sensor nodes, it is possible to obtain real-time data about any physical phenomena, that were difficult or impossible to obtain in conventional ways of networking. Hence a typical WSNs application is combined different software and hardware module to sense the physical data, processing data and communicates those data to a remote site or base station through multiple hops or nodes. Sometimes, instead of sending the raw data to the nodes responsible for the fusion, they use their processing abilities to locally carry out simple computations and transmit only the required and partially processed data [3].

In WSNs, energy is  one of the major issue, which need to be carefully consumed by the sensor nodes to maximize the network lifetime. Hence, most of the current researchers are engaged to devise a way to minimize the energy consumption in the network to maximize the network lifetime. Generally the sensor nodes inorder are powered by small batteries which are incapable to power for a long period. Typically, the sensors nodes are deployed in a left unattended area. So, it is quite difficult to replace the battery frequently and even sometimes not possible. Therefore, prolonging the network lifetime is an important optimization goal in this aspect. The secret to reduce energy consumption lies in power aware designing of each layer of the system and number of message transmission for data packet. As a consequence, the reduction of number of message transmission can reflects in energy uaseg of the nodes. The WSNs suffer from several constraints and challenges; some of those are: limited onboard memory and limited processing capability of the sensor nodes, limited communication bandwidth, frequent death of sensor nodes, link failure, mobility in sensor, heavy traffic through some particular nodes etc. In a multihop ad hoc sensor network, each node plays the twins role of data originator and data router. The failure of few nodes may lead to topological changes and hence require rerouting of the data packets.

To resolve the issue of excess energy consumption, many energy-efficient protocols have been proposed to reduce energy consumption such as S-MAC [4], LEACH [5], CLEEP [6], SPIN [7] and etc. Many MAC protocols have been focused to avoid the collision between two nodes and minimizing the idle listening; but ignore the route discovery and maintenance issues. Hence, the literal challenging task, do not fulfill, which in turn results inefficient energy consumption.

This paper introduces a novel adaptive approach to find an optimal routing path from source to sink when the sensor nodes are deployed randomly deployed in a restricted service area with single sink. This also aggregate the data in intermediate node to reduce the duplicate data. Data centric protocols more focus on data rather than the address of the destination. Here our approach focus on both data as well as the destination address. The proposed protocol is hierarchical in nature. The work looks at a unique approach for a scalable and energy efficient solution. The proposed protocol has been evaluated by performance analysis with the existing protocol FLOODING and SPIN.

The rest of the paper is organized as follows: Section II mentions some related works. Section III, describes the proposed protocol in detail. Section IV presents the simulation results and Section V summarizes the paper with a conclusion of the work.

\section{Related Work}
Limited energy resources of sensor nodes endlessly insist the researchers to work for new algorithms and schemes to prolong the lifetime of the network. Thus energy efficient strategies have been focused in each layer of protocol design in network. Although these protocols have achieved a good level of performance in their respective layers, but they usually ignores the impact of other layers on their performance, which in turns inefficient in its integrated implementation.

One indicated approach to adaptively adjust the transmission power to an appropriate level for generating signal strength, just enough to reach the next hop destination, is to control the power consumption rate of a sensor node and thus to reduce the collision probability [8]. The S-MAC [4] protocol used a low power RTS-CTS protocol periodically sleeps, wakes up, listens to the channel, and then returns to sleep [1], hence adresses collision problem in network. In MAC layer, the major sources of energy waste in wireless sensor networks like collision, overhearing, control-packet overhead, idle listening and over emitting should be minimize to prevent the energy wastes [2]. The AC-MAC/DPM [9] introduced Dynamic Power Management (DPM) mechanism into ACMAC to reduce the energy consumption due to the transceiver state switches between idle and sleep. The Sift [10] addresses the issue of when multiple nodes in the same neighborhood all sense an event they need to transmit information about and a subset of the nodes that observe the same event report it. P-MAC adaptively determines the sleep-wake up schedules for a node based on its own traffic, and the traffic patterns of its neighbors [5]. Routing in WSNs is very challenging issue due to inherent characteristics that distinguish the WSNs from other wireless networks like mobile ad hoc networks or cellular networks [8]. The sensor network adopts two basic schemes for energy savings in network layer such as [3]: Power-Aware Routing   and Maximum Lifetime Routing. The main aim at the network layer is to find the route to transmit data from sensor nodes to the sink in an energy-efficient and reliable manner in order to maximally extend the lifetime of the network. The SPIN [7], an adaptive protocols that disseminate all the information from each node to every node in the sensor network. The SPIN family of protocols uses data negotiation and resource-adaptive algorithms. SPIN protocol delivers 60\% more data with consuming same amount of energy as flooding [7].

In recent years, many protocols also have been designed to provide energy-efficient cross layer mechanism such as Energy Aware Adaptive Low Power Listening (EA-ALPL), Cross-Layer Optimization, and Cross-Layer Scheduling. In comparison, our approach promotes a better performance as compare when with flooding and LEACH [5] protocols.

%
%
\begin{figure}[t]
\centering
\includegraphics[width=3in]{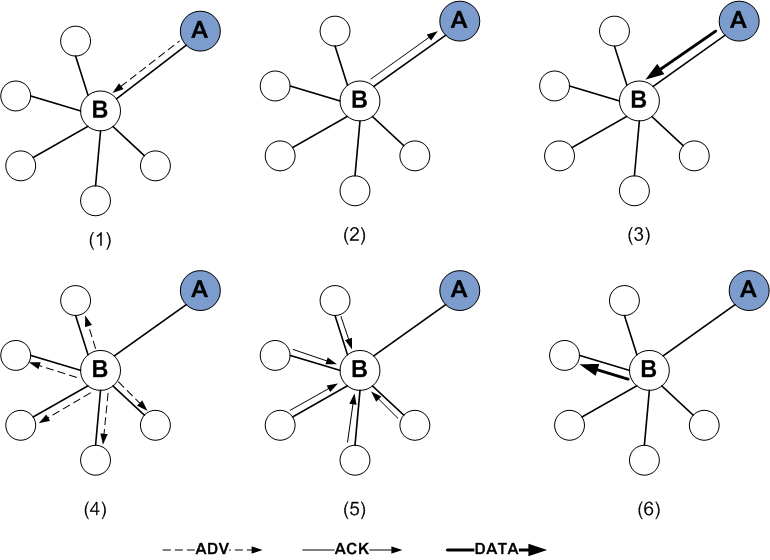}
\caption{DRUG protocol working model. Node A starts by advertising its data to node B (1). Node B responds by sending a request to node A (2). After receiving the requested data (3), node B then sends out advertisements to its neighbors (4), who in turn send requests back to B (5-6).}
\label{fig_sim}
\end{figure}

\section{MAC Layer Related Properties of WSNs}
\subsection{Reasons of Energy Wastes}
The sensor nodes are generally disposed when they are out of battery. So Max-
imizing the network lifetime is a major question of sensor network design.The
implemented MAC protocol should reduce the potential energy wastes.The com-
mon reasons of energy wastes found are:
\begin{itemize}
\item Idle Listening: It is one of the major sources of energy waste.When a
node listen to an idle channel to receive possible traffic,It is called idle
listening.
\item Collision: When a receiver node receives more than one packet at the
same time, these packets are called collided packets even when they co-
incide partially. All packets that cause the collision have to be discarded
and the re-transmissions of these packets are required which increase the
energy consumption. Although some packets could be recovered by a cap-
ture effect, a number of requirements have to be achieved for its success.
\item Overhearing: Overhearing happens a node receives packets that are des-
tined to other nodes.
\item Control packet overhead: The third energy waste occurs as a result of
control packet overhead. Minimal number of control packets should be
used to make a data transmission.
\item Overemitting: It is caused by the transmission of a message when the
destination node is not ready.
\end{itemize}
\subsection{Communication Patterns in WSNs}
Three types of communication patterns are seen in wireless sensor networks
broadcast, convergecast, and local gossip.

Broadcast type of communication pattern is generally used by a base station
(sink) to transmit some information to all sensor nodes of the network. The
event triggered sensors communicate with each other locally. This kind of com-
munication pattern is called local gossip, where a sensorsends a message to
its neighboring nodes within a range.The sensors that detect the event, then,
need to send what they perceive to the information center. That communica-
tion pattern is called convergecast, where a group of sensors communicate to
a specific sensor. The common features of communication patterns are :
\begin{itemize}
\item Little activity in lengthy period
\item Intensive traffic in short time
\item Highly correlated traffic
\item End to end flows are required to be fair
\end{itemize}
\section{Energy Consumption Model}
In WSNs, the operation time of individual sensor node greatly influence the whole lifetime of the network. Therefore, a model, which defines the amount of power consumed in each action of a sensor node, influences the lifetime of networks to a great degree. In proposed work, we assume a model where the radio dissipates \(E_{elec}\) = 50nJ/bit to run the transmitter or receiver circuitry and \(\varepsilon_{amp}\) = 100pJ/bit/m\(^2\)
for the transmit amplifier to achieve an acceptable \(E_{b}\)/No [7, 13]. The power needed to transmit $k$ bits of data over a distance $d$ is:
\begin{equation}
E_{tx} = E_{elec} k + \varepsilon_{amp} kd^2
\end{equation}
And the power needed to receive $k$ bits of data is:
\begin{equation}
E_{rx} = E_{elec} k
\end{equation}

Where, $d$ is the distance between the source and sink. Using a direct communication protocol, each sensor sends its data directly to the base station. If the base station is far away from the nodes, direct communication will require a large amount of transmit power from each node. This will quickly drain the battery of the nodes and in turn reduce the network lifetime. Nodes route their packets to the base station through intermediate nodes. Thus nodes act as routers for other nodes in addition to sense the environment. The existing routing protocols consider the energy of the transmitter and neglect the energy dissipation of the receiver in determining the routes in equation 2.

Depending on the relative costs of the transmit amplifier and the radio electronics, the total energy expended in the system might be greater in multi-hop transmission than direct transmission to the base station.

Assume that there are $n$ numbers of intermediate nodes to reach at the destination and also each adjacency nodes are differentiated with distance $r$ between them. So, the total distance between source to sink is $nr$. If we consider the energy expenditure at each node during transmitting a single $k$-bit message from source node $N$ to base station. A node located with a distance from the base station using the direct communication approach is in equations 1 and 2, then from equation 1.
\begin{equation}
\begin{split}
E_{direct} & =  E_{tx}(k, d = n \ast r)\\
& = E_{elec} \ast k + \varepsilon_{amp} \ast k \ast (nr)^2\\
& = k (E_{elec} + \varepsilon_{amp} n^2 r^2)
\end{split}
\end{equation}

Packet passes through the $n$ intermediate nodes to reach at the destinations means it required $n$ times transmit and $n-1$ time receive. From equation (2)
\begin{equation}
E_{rx} = (n-1) E_{elec} k
\end{equation}
So total energy conservation to reach at the destination is
\begin{equation}
\begin{split}
E & = n (E_{elec} \ast k + \varepsilon_{amp} \ast k \ast r^2) + (n-1) E_{rx}\\
& = E_{elc} * k * n + \varepsilon_{amp} \ast k \ast n \ast r^2 + (n-1) E_{elec} k\\
& = k ((2n-1) E_{elec} + \varepsilon_{amp} nr^2)
\end{split}
\end{equation}
In the direct communication with base station the energy conservation is
\begin{equation}
\begin{split}
E & = E_{tx} + E_{rx}\\
& = E_{elec} k + \varepsilon_{amp} kd^2 + E_{elec} k\\
& = E_{elec} k + \varepsilon_{amp} kr^2 + E_{elec} k\\
& = k (2 E_{elec} + \varepsilon_{amp} r^2)
\end{split}
\end{equation}
From the above equations the total energy at $n$ hop distance from the source to sink is defined in equation 5 and for single hop communication in equation 6.

\section{DRUG Protocol Overview}
In this section, we have presented the new data centric protocol i.e. DRUG (Datacentric RoUtinG) and described the working model. This protocol resets upon two basic ideas. First to operate efficiently and to conserve energy, sensor applications need to communicate with each other about the data that they already have and the data they still need to obtain. Exchanging sensor data may be an expensive network operation, but exchanging data about sensor data (say metadata) need not be. Second, nodes in a network must monitor and adapt to changes in their own energy resources to extend the operating lifetime of the system. This is the main feature of the DRUG protocol. 
\subsection{Application-level control}
Our design of the DRUG protocols is motivated in part by the principle of Application Level Framing (ALF) [2]. With ALF, network protocols must choose broadcast units that are significant to applications, i.e., packetization is best done in terms of Application Data Units (ADUs). One of the important components of ALF-based protocols is the common data naming between the transmission protocol and application, which we follow in the design of our meta-data [1]. We take ALF-like ideas one step further by arguing that routing decisions are also best made in application-controlled and application-specific ways, using knowledge of not just network topology but application data layout and the state of resources at each node. We believe that such integrated approaches to naming and routing are attractive to a large range of network situations, especially in mobile and wireless net-works of devices and sensors. 
\subsection{Meta-data}
Sensors use meta-data to represent the description about data collected by the sensor. If $x$ is the meta-data description of sensor data X, then the size of $x$ in bytes must be shorter than the size of $X$ [1]. If different pieces of actual data are distinguishable, then their corresponding meta-data also be distinguishable. Likewise, two pieces of same data should share the same meta-data representation.

It does not specify a format for meta-data; this for-mat is application-specific. For example, sensors that cover disjoint geographic regions may simply use their own unique IDs as meta-data [3]. The meta-data $x$ would then stand for ``all the data gathered by sensor $x$". By contrast, a camera sensor might use ($x$, $y$, $\phi$) as meta-data, where($x$, $y$) is a geographic coordinate and $\phi$ is an orientation. DRUG applications must take care to define a meta-data format for representing data that takes into account the costs of storing, retrieving, and managing the meta-data [1]. Finally, because each application's meta-data format may be different, DRUG relies on each application to interpret and synthesize its own meta-data.
\subsection{DRUG messages}
In our proposed desinged, the DRUG protocol uses three types of messages to communicate between different nodes as shown in Fig. 1, such as:
\begin{itemize}
\item ADV: new data advertisement. When a sensor node has data to share, it can advertise this fact by transmitting an ADV message containing meta-data.
\item ACK: request for data. A SPIN node sends an ACK message when it wishes to receive data.
\item DATA: data message. DATA messages contain actual sensor data with a meta-data header.
\end{itemize}

ADV and ACK messages contain only meta-data. In networks where the cost of sending and receiving a message is largely determined by the message’s size, ADV and ACK messages will therefore be cheaper to transmit and receive than their corresponding DATA messages.

\subsection{DRUG resource management}
The applications of proposed DRUG protocols are resource-aware, resource-adaptive and data centric like SPIN properties. They calculate the cost in terms of energy, of performing computations and sending and receiving data over the network. With this information, nodes can make informed decisions about using their resources effectively. Both the protocols follow the data centric method. SPIN does not specify a particular energy management policy for its protocols. Rather, it specifies an interface that applications can use to probe their available resources [1]. DRUG protocol transmits data with three step process like SPIN but it transmit with unicast whereas SPIN transmit with multicast. Our proposed method takes care about the data centric as well as the address of sink node.
\begin{algorithm}[t]
\caption{Network Initialization}
Each node associates with one largest integer value i.e $\infty$ \newline
V[X]= associate value with the node X \newline
Q= Queue 
\hrule
\begin{algorithmic}[1]
\STATE $V[Sink] \leftarrow 0$
\STATE $Q \leftarrow Sink$ \text Assign the Sink to the queue
\STATE $X \leftarrow Dequeue(Q)$
\STATE \text Node(X) broadcast the ADV packet containing the associate integer value.
\STATE \text Node(Y) receive the ADV packet and check.
\IF {$V[Y] \geq V[X]$}
\STATE $V[Y] = V[X] + 1$
\STATE $Q \leftarrow Y$ \text Node Y En-queue in to the queue.
\ENDIF
\STATE \text Continue to Step 3 till Queue (Q) is empty.
\IF{$|Q| \neq 0$}
\STATE \text Go to Step 3.
\ELSE
\STATE $Exit$ \text Network initialization completed.
\ENDIF
\end{algorithmic}
\end{algorithm}

\begin{algorithm}[t]
\caption{Data transmission from event node}
id(X) = ID of node X \newline
V[X] = value associated with node X \newline
En(X) = Energy status at node X \newline
En = Minimum energy level to participate in data transmission
\hrule
\begin{algorithmic}[1]
\STATE $B\_cast ADV(id(X),V[X])$
\STATE \text Node(Y) receive the ADV packet and check
\nonumber \IF {$V[Y] < V[X]$ AND $En(Y) \geq En$}
\nonumber \STATE $ACK(Y) \rightarrow X$
\nonumber \ENDIF
\STATE $DATA(X) \rightarrow Y$
\STATE \text Node(Y), B\_cast ADV
\STATE \text Go to Step 3 till DATA packet reach at the sink
\end{algorithmic}
\end{algorithm}

\section{Simulation Results}
In this section we evaluate the performance result of the proposed protocol and compared with the existing SPIN protocol. The protocol has been analyzed and evaluated in NS2 simulator. Here we compared with SPIN protocol because, it is an standard protocol for the data centric transmission. To perform the test, we have taken 100 random sensor nodes in the 1000x1000 meter area.

\begin{figure*}
  \centering
  \mbox
  {
    \subfigure[First node dies in the network. \label{subfigure label}]{\includegraphics[scale=.45]{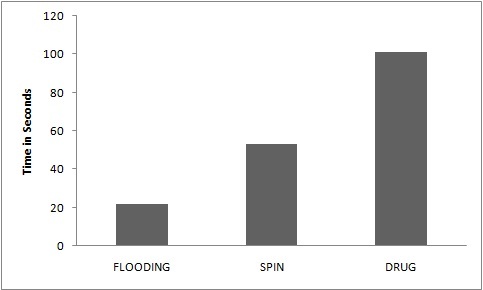}}\quad \label{fig:14}
    \subfigure[Delivery Ratio (\%) Vs Time (Sec.). \label{subfigure label}]{\includegraphics[scale=.32]{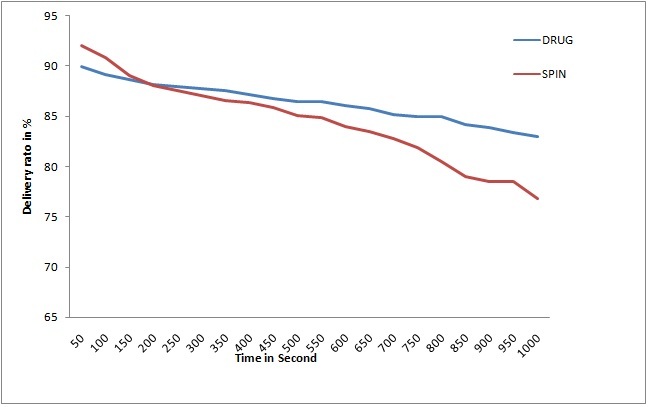}}\quad \label{fig:15}
    \subfigure[Ratidual Energy (Joule) Vs Time (Sec.). \label{subfigure label}]{\includegraphics[scale=.28]{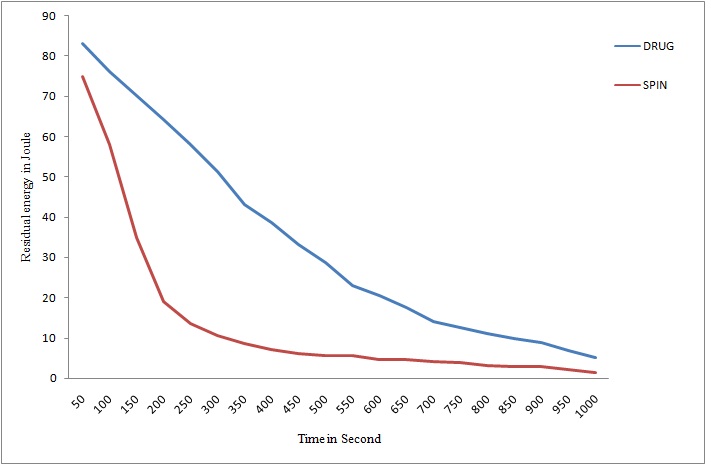}} \label{fig:16}
  }
  \caption{Performance comparision of our proposed DRUG protocol.}
  \label{fig:Fig5}
\end{figure*}

The Figure 2 (a) shows the first node dies in the network with considering various technologies. In flooding the first node dies very quickly in the considering scenario because it floods the data packets to entire network in-order to deliver the data packets. Comparatively flooding, SPIN saves more energy and sends the data to the destination. It sends the data after establishes the path and follow the same path until it breaks. In this way the node dies slowly. In the MSWSN model more energy saves and all nodes of the network are alive for long period of time.
 
In the Fig. 2 (b) we have shown the delivery ratio between the SPIN and DRUG routing protocol. Initially in SPIN delivery ratio is higher than the SPIN because of their flooding nature of data transmission. As soon as node dies, delivery ratio decreases. In our proposed model delivery ratio performance is better than SIPN protocol.

In Fig. 2 (c), shows at initially of simulation residual energy of the network is very less in SPIN. The reason behind the drastic decrement of residual energy of the network is the broadcasting nature of the node. A large number of nodes die because of this resion and further the network becomes disconnected. In our proposed protocol DRUG data dissemination towards a single mode instead of multiple node and transmits with negotiation based in order to reach at the destination. When a node reduces its energy below threshold level, it is not going to participate in data transmission unlike SPIN property. So that the decrement in the residual energy become almost constant in the rest of the experiment. So the decrement of residue energy is smooth. 

\section{Conclusion}
In this paper, a new energy-efficient data-centric routing protocol for wireless sensor networks is presented, named DRUG. The protocol is motivated in part by the principle of application level framing. We take ALF-like ideas one step further by arguing that routing decisions are also best made in application-controlled and application-specific ways, using knowledge of not just network topology but application data layout and the state of resources at each node. The results of analytical simulation experiment show that DRUG conserves more energy and leads to the better system performance. 






%

\end{document}